\documentclass[structabstract]{aa}
\usepackage{graphicx}
\usepackage{natbib}
\bibpunct{(}{)}{,}{a}{}{,}
\usepackage{txfonts}
\begin{document}
   \title{Mesogranular structure in a hydrodynamical simulation}

   %\subtitle{III. Three-dimensional hydrodynamic simulation.}

   \author{{\L}. Matloch\inst{1} \and
          R. Cameron\inst{1} \and
          S. Shelyag\inst{2} \and
          D. Schmitt\inst{1} \and
          M. Sch\"{u}ssler\inst{1}
          }

   \institute{Max-Planck-Institut f\"{u}r Sonnensystemforschung, Max-Planck-Strasse 2, 37191 Katlenburg-Lindau, Germany\\
              \email{matloch@mps.mpg.de}
         \and
             Astrophysics Research Center, Queen's University Belfast, Belfast BT7 1NN, Northern Ireland, UK
             }

   \date{Received; accepted }

%\abstract{}{}{}{}{}
% 5 {} token are mandatory

  \abstract
   {}
   {We analyse mesogranular flow patterns in a three-dimensional hydrodynamical simulation of solar surface convection in order to determine its
   characteristics.}
   {We calculate divergence maps from horizontal velocities obtained with the Local Correlation Tracking (LCT) method. Mesogranules are identified
   as patches of positive velocity divergence. We track the mesogranules to obtain their size and lifetime distributions. We vary the analysis
   parameters to verify if the pattern has characteristic scales.}
   {The characteristics of the resulting flow patterns depend on the averaging time and length used in the analysis.}
   {We conclude that the mesogranular patterns do not exhibit intrinsic length and time scales.}


   \maketitle

\section{Introduction}
Flow patterns on scales between granulation and supergranulation, are found in both observations and hydrodynamical simulations. They appear in
divergence maps of horizontal flows inferred by local correlation tracking (LCT) of solar granules and magnetic flux concentrations
\citep{November,Simon SA,Roudier1998,PlonerA,Cardena,Cattaneo2001}. In our previous work (Matloch et al. 2009) we studied simplified one- and
two-dimensional granulation models in order to investigate the origin of such mesogranular flow patterns. We showed that patterns very
similar to those observed emerge in such models and that they can be attributed to the local interactions between granules together with
the spatial and temporal averaging used to analyse the data. In this paper, we investigate mesogranular patterns in a hydrodynamical
simulation of solar surface convection. We apply the same definitions and analysis methods as were used for the two-dimensional
model to allow a direct comparison of the results. If mesogranular flow patterns result from a self-arrangement of granules, as suggested
by our previous models, the pattern should be present in the numerical simulation as well.

In Section~2 we briefly describe the simulation setup. Section~3 contains a comparison of the LCT velocities with the actual
plasma velocities. In Section~4 we present the analysis methods and results, while Section~5 contains the
conclusions.
\section{Simulation}
\begin{figure}[t]
\begin{center}
\includegraphics[width=90mm]{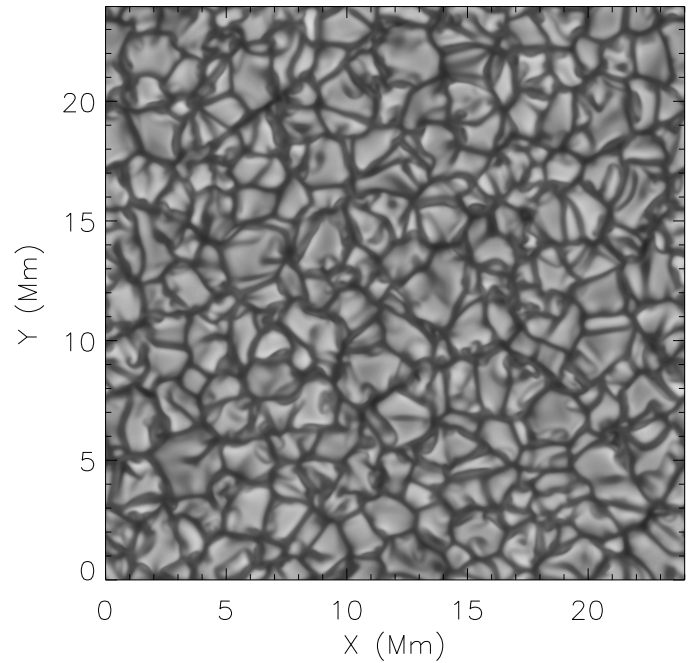}
\end{center}
\caption{Bolometric brightness distribution from the simulation. Plasma flows upward in the bright cell interiors
(granules) and back into the interior in the darker intergranular lanes.} \label{figmuram}
\end{figure}

The MURaM code (V\"{o}gler 2003, V\"{o}gler et al. 2005) treats the equations of compressible (magneto-) hydrodynamics,
incorporating radiative transfer and partial ionization effects in local thermal equilibrium. The simulation run analysed
here has a domain size of $24\times24\times2.3$ $\mathrm{Mm}^3$, with periodic horizontal boundary conditions. The $\tau=1$
level is located roughly $600$ km below the top of the simulation box. The grid resolution is $20.8$ km in the horizontal and
$14$ km in vertical direction. The magnetic field is set to zero. The bottom boundary is open and allows for mass flow. The
top boundary is closed, with vanishing horizontal viscous stress. The total length of the simulation run is $11$ hours.
Figure~\ref{figmuram} shows a brightness snapshot from the simulation.
\begin{figure*}[t]
   \centering
   \includegraphics[width=90mm]{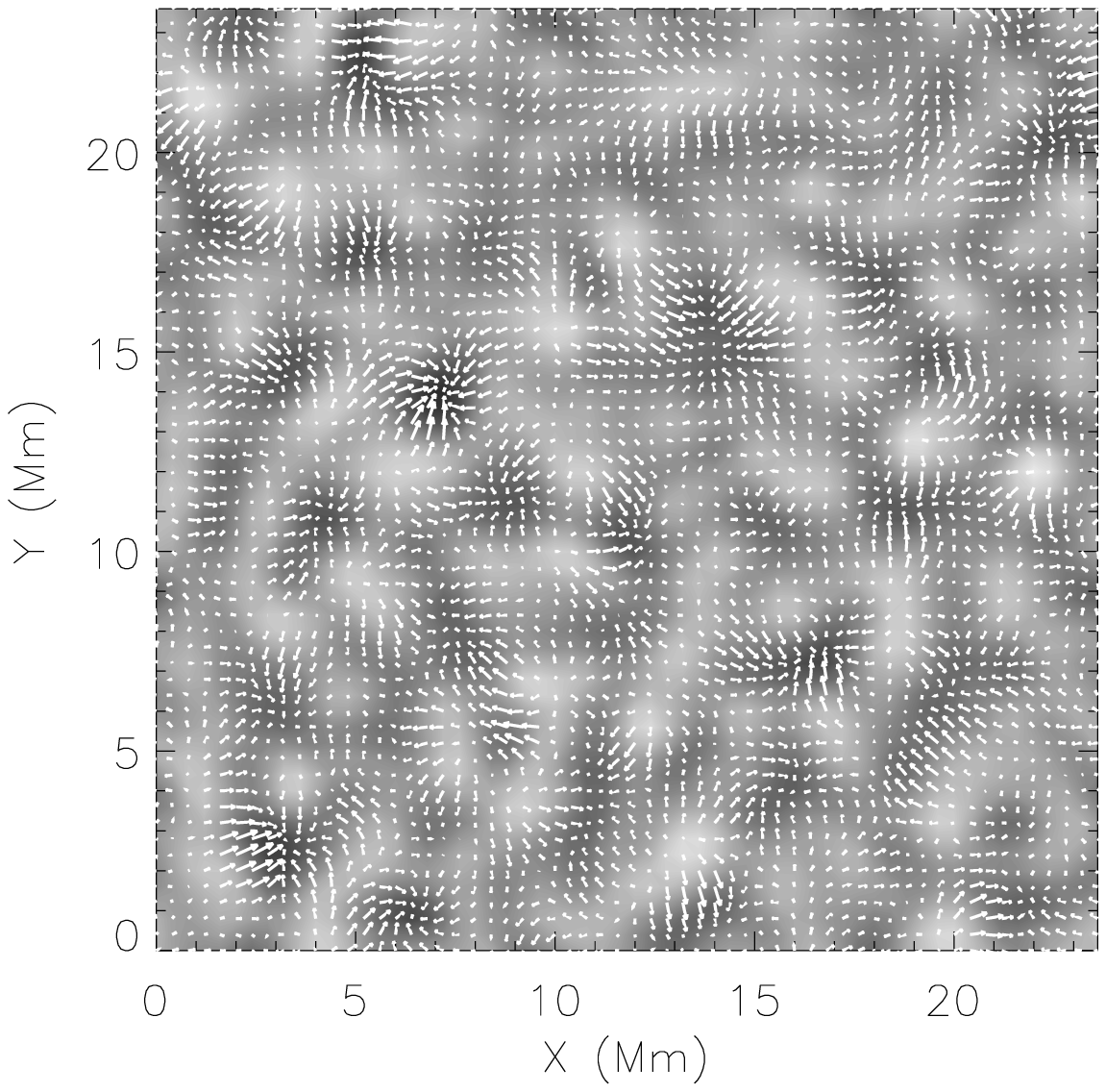}
   \includegraphics[width=90mm]{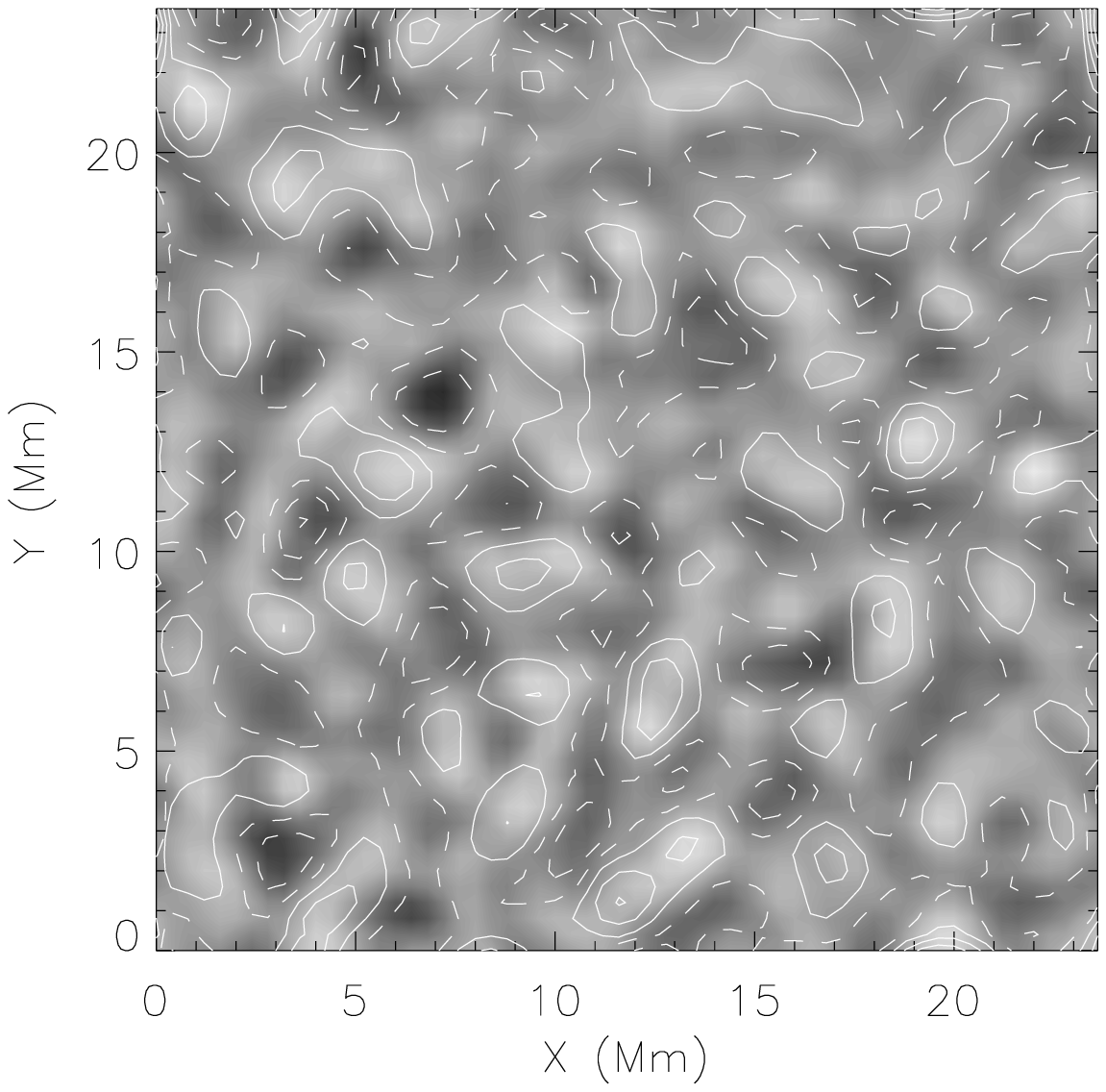}
   \includegraphics[width=90mm]{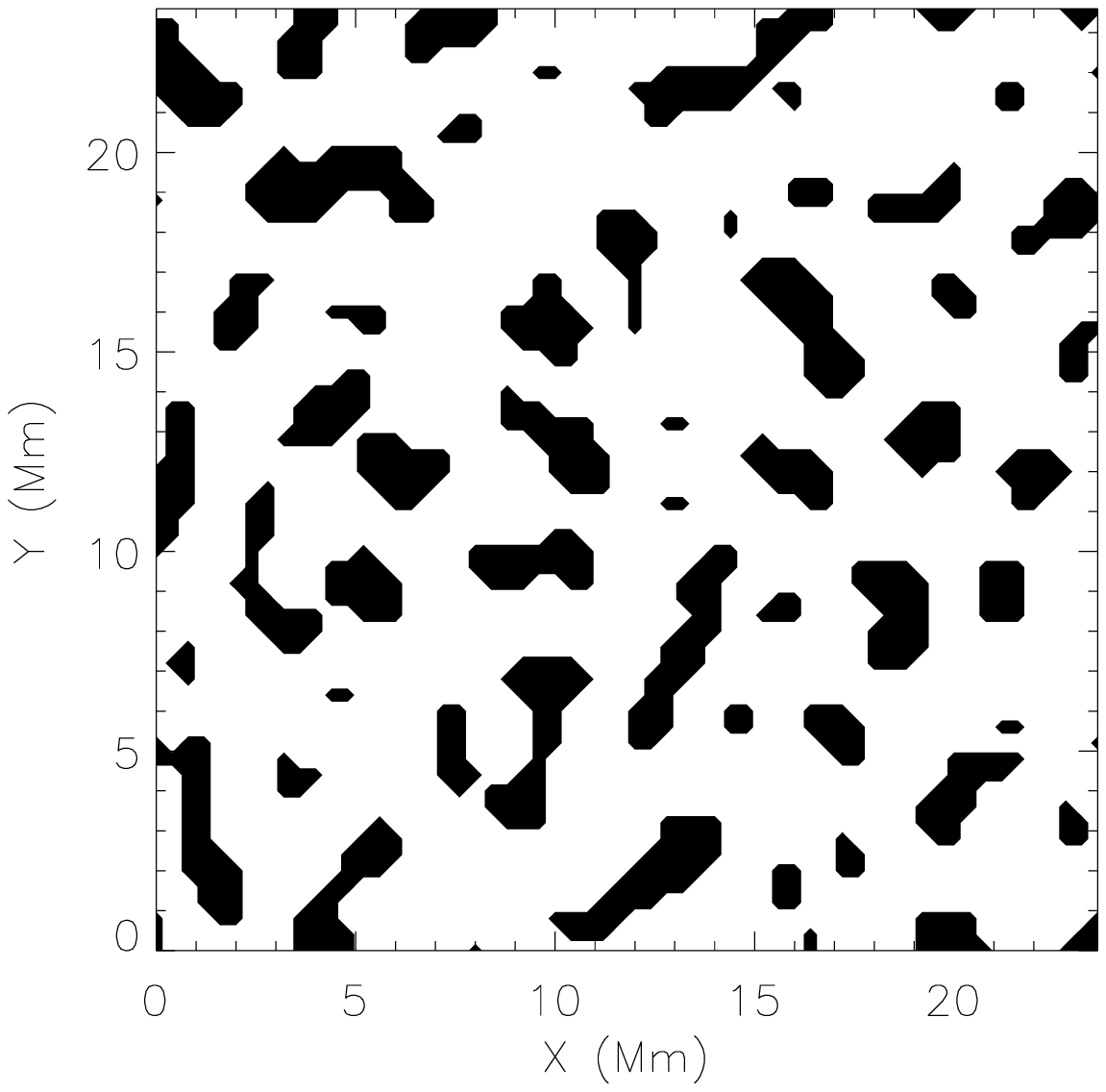}
   \includegraphics[width=90mm]{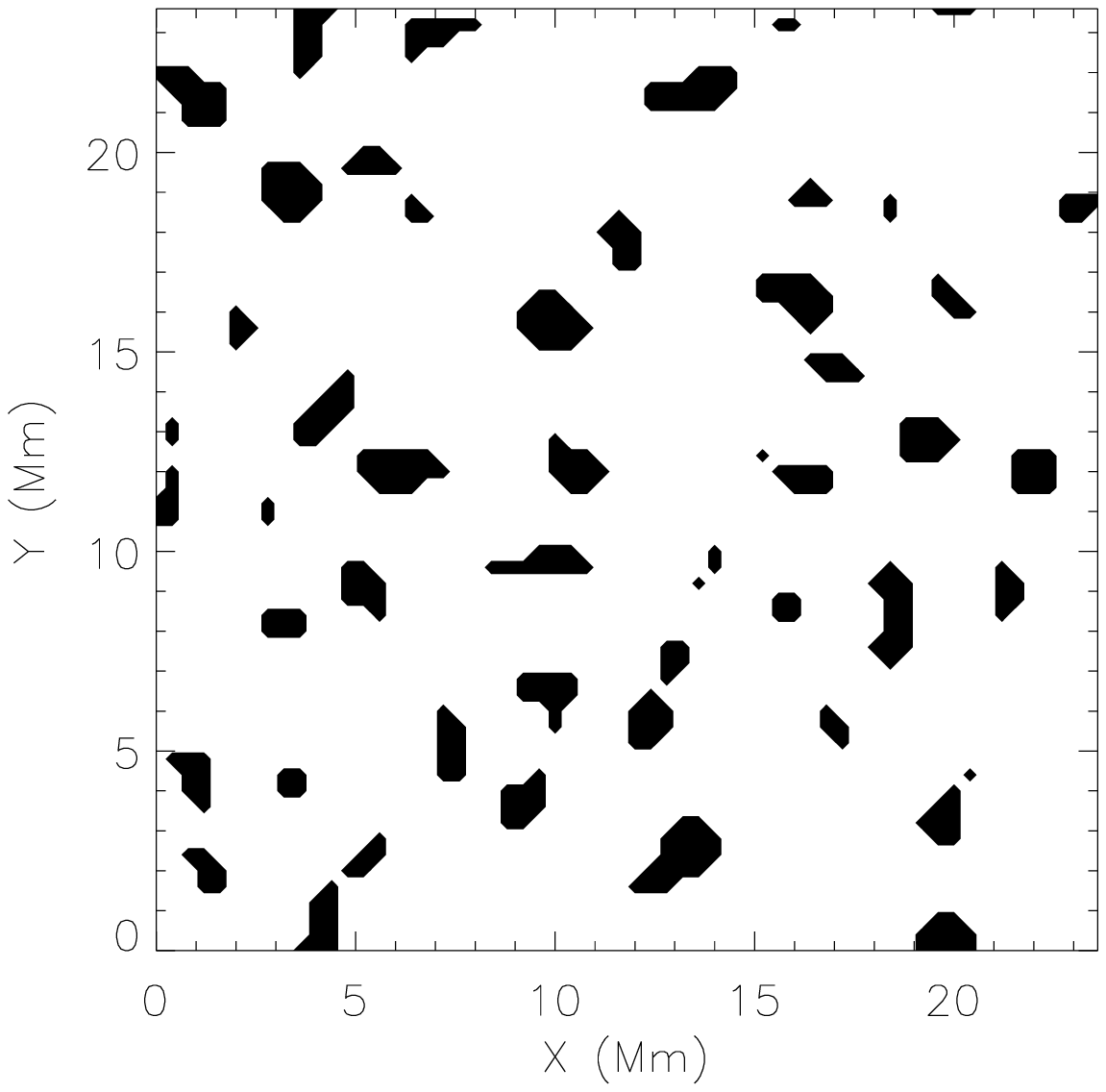}
         \caption{Top left: LCT horizontal velocity divergence field (grey-scale, bright: positive, dark: negative) and the
velocity arrows, temporally averaged over $60$ minutes. The longest arrows correspond to velocities of $\sim540$
$\mathrm{ms^-1}$. Top right : the same LCT horizontal velocity divergence field, overplotted with the contours of the
corresponding divergence of the actual flow velocity averaged spatially over the LCT window size. Solid lines indicate
positive divergence, dashed negative divergence. Bottom left: corresponding velocity divergence patches (mesogranules) lying
above the $0.5\Lambda$ threshold (see explanation in Sect. 4). Bottom right: the same but for the $0.7\Lambda$ threshold.}
         \label{velovect}
   \end{figure*}
\section{Local Correlation Tracking}
In observations and simulations mesogranules are often identified with areas of positive horizontal velocity divergence
\citep{Roudier1998,Cattaneo2001,Leitzinger2005}. In observations the horizontal velocity field is obtained with a LCT
algorithm, which tracks intensity patterns on the surface. Figure~\ref{velovect} (top left panel) shows the divergence of
horizontal velocity obtained from the simulation with the LCT method, averaged over $60$ min, along with the corresponding
velocity arrows. We use the LCT algorithm described by \citet{Welsh2004}, with the tracking window being a Gaussian with FWHM
of $1$ Mm. Using data from the MURaM simulation, we find that the LCT velocities are roughly between $0.5-0.7$ in magnitude
as compared to the actual plasma velocities around continuum optical depth unity (see also Rieutord et al. 2001, Georgobiani
et al. 2007). The top right panel of Fig.~\ref{velovect} shows the LCT velocity divergence field overplotted with contours of
the divergence of the actual velocity from the simulation, averaged spatially over the LCT window size. Both velocity fields
are temporally averaged over $60$ minutes and have a cross correlation coefficient $0.75$. The mean correlation value between
the divergences of the LCT and actual plasma velocities for the whole dataset equals $0.73$.
   \begin{figure}[t]
   \centering
   \includegraphics[width=90mm]{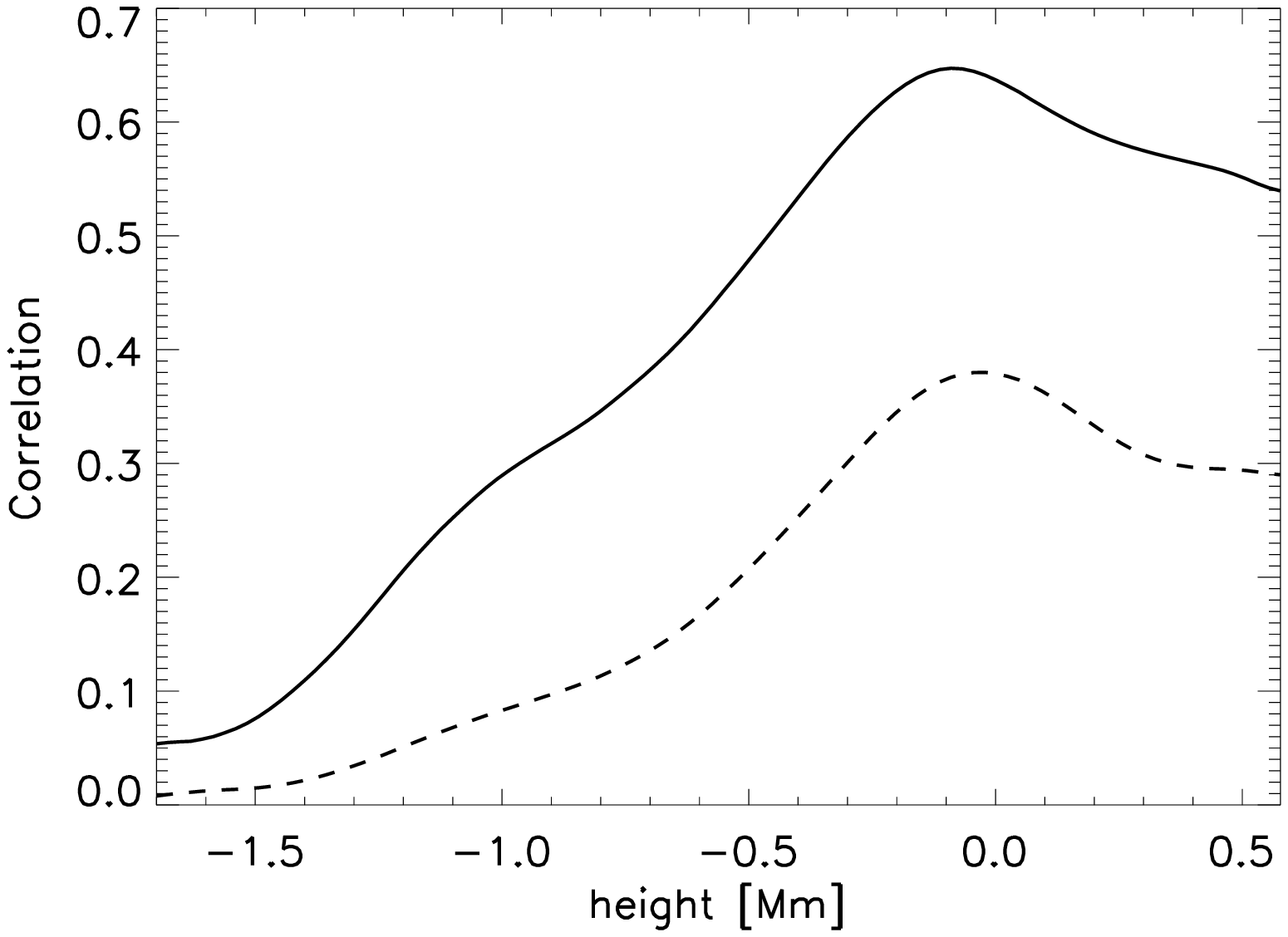}
   \includegraphics[width=90mm]{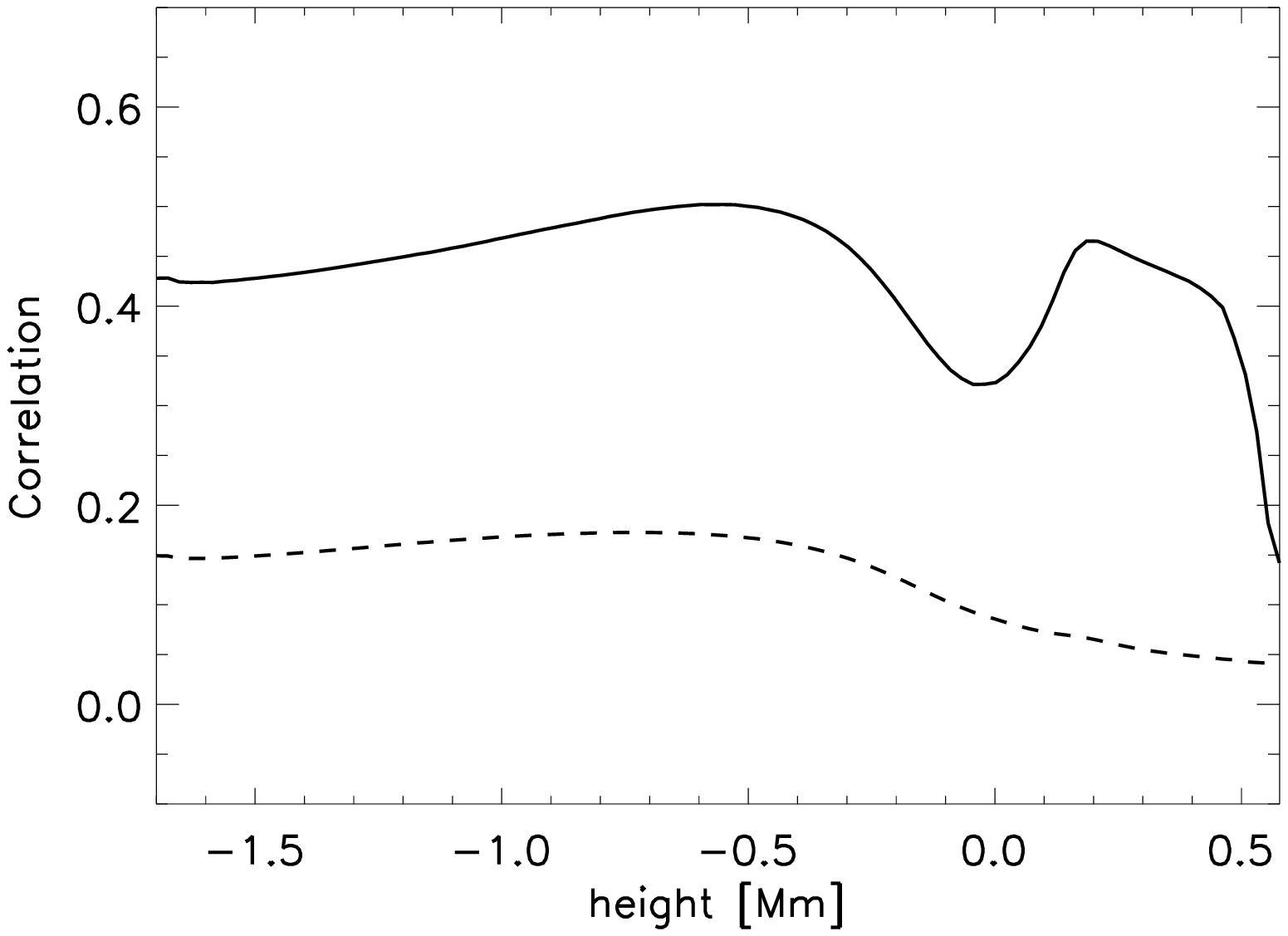}
         \caption{Top: correlation between the LCT velocities and the horizontal plasma velocities coming from different
depths of the simulation box. Bottom: Correlation between the LCT velocity divergence and vertical plasma velocities coming
from different depths of the simulation box. The solid line represents the correlation coefficient as a function of depth for
the $47$-min average of the data. The dashed line shows the correlation coefficients calculated individually for each of the
$81$ velocity snapshots within the $47$ minutes, and then averaged.}
         \label{figcoor}
   \end{figure}
   \begin{figure}[t]
\begin{center}
\includegraphics[width=78mm]{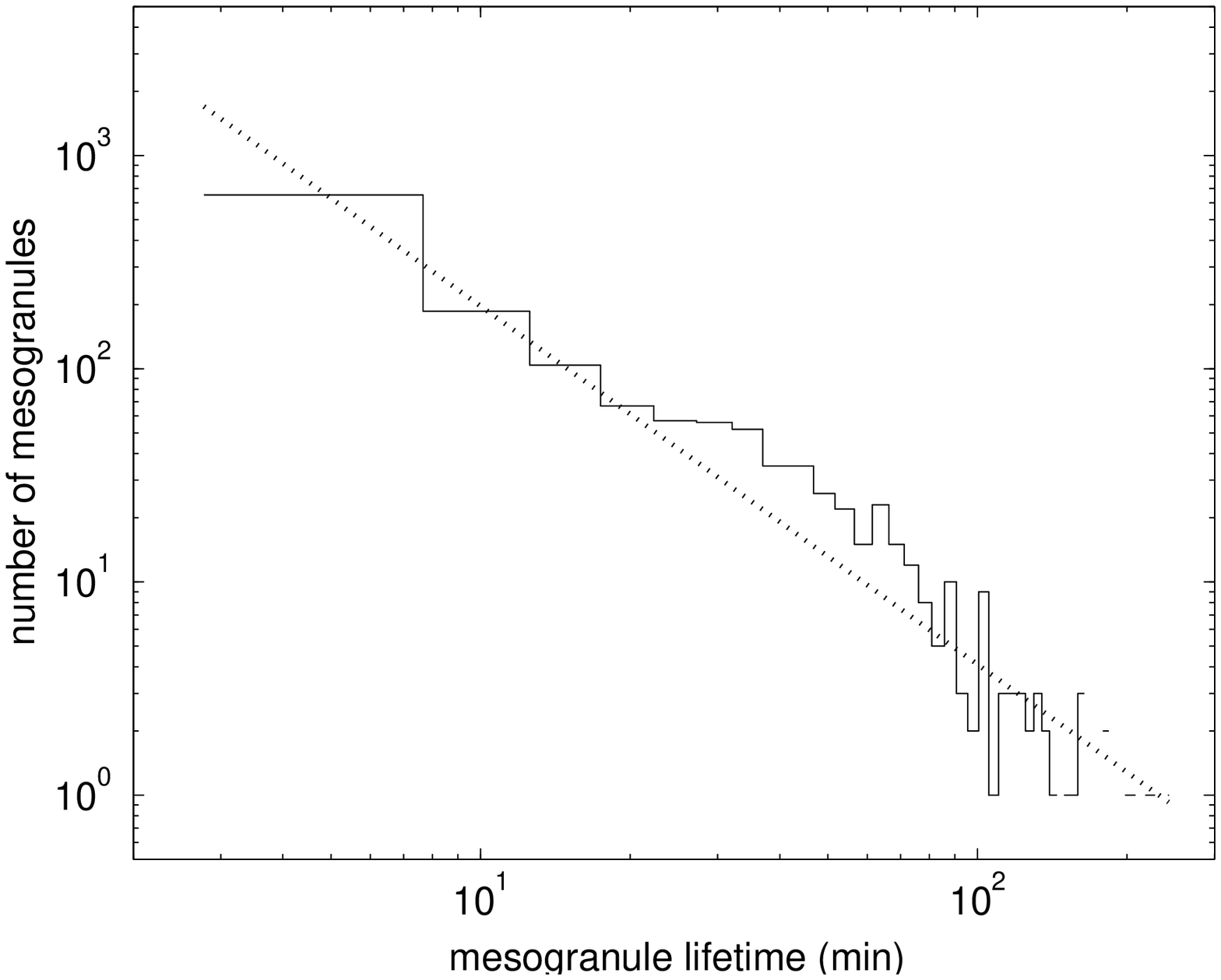} %height=45mm,
\includegraphics[width=78mm]{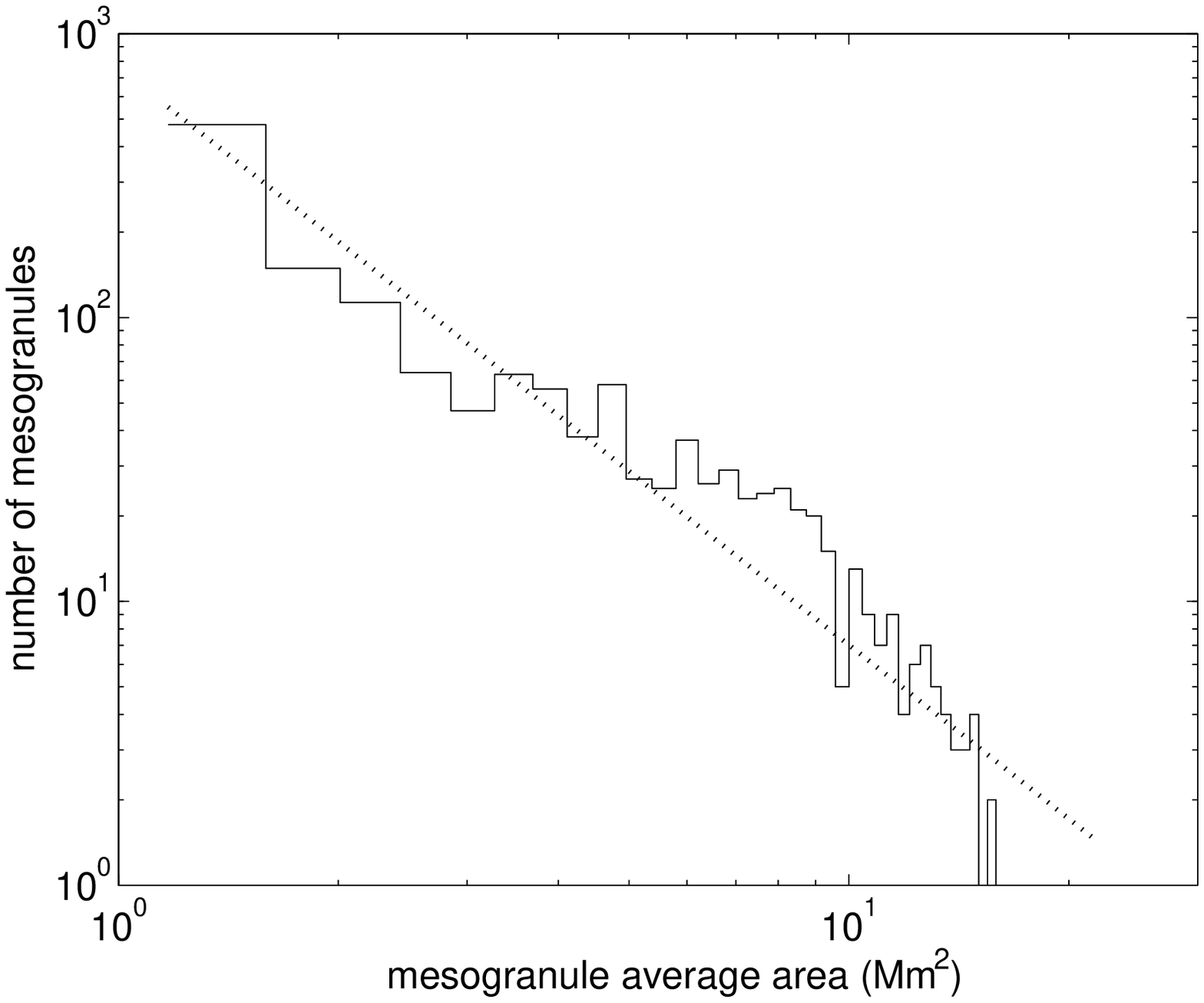}
\includegraphics[width=85mm]{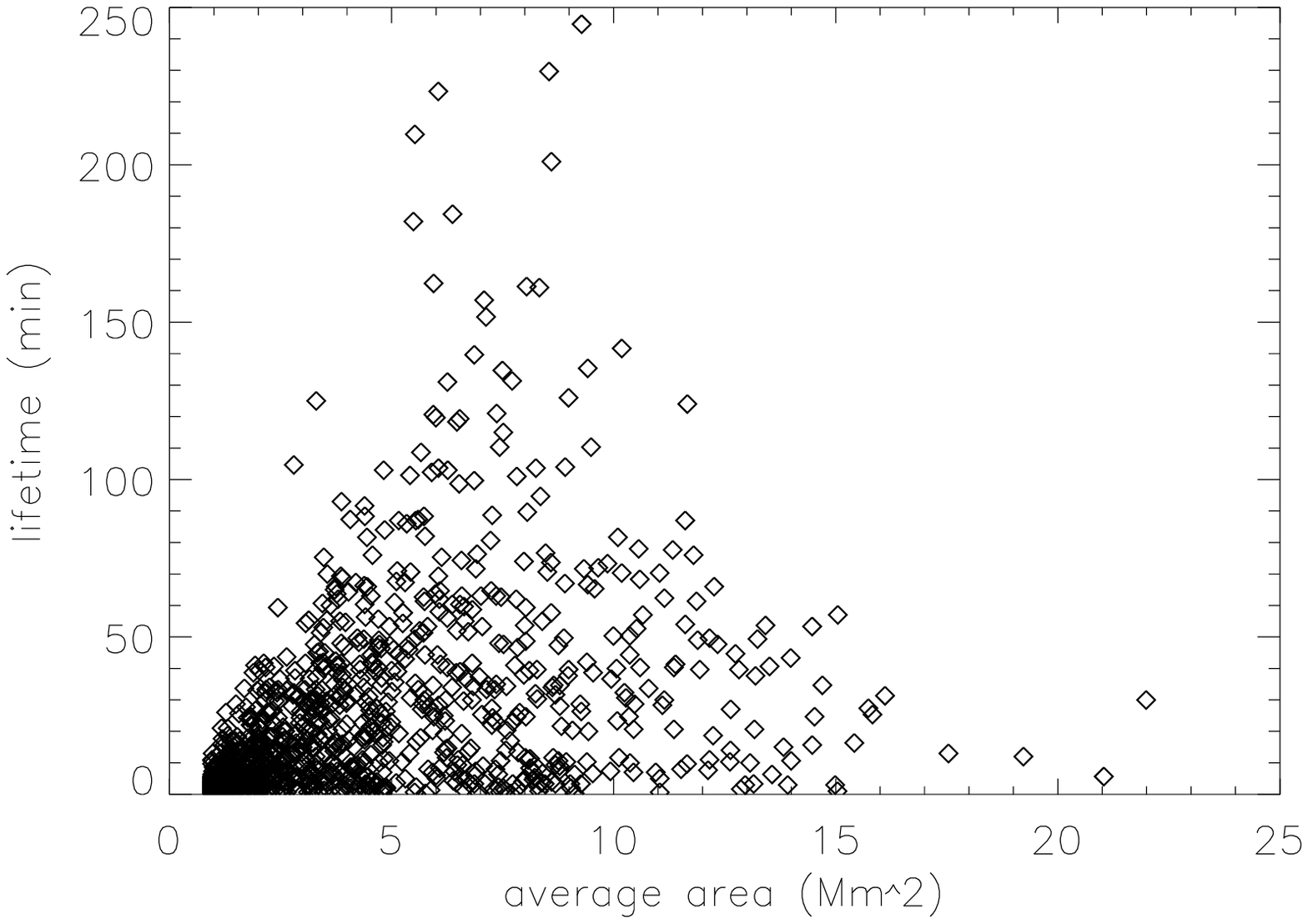}
\end{center}
\vspace{8mm} \caption{Mesogranule lifetime (top) and size histograms (middle), together with a scatter plot of area versus
lifetime (bottom). The threshold level and averaging time are $0.5\Lambda$ and $t_{a}=60$ min., respectively. Dotted lines
represent power-law fits with exponents $-1.2$ for the lifetime and $-2.4$ for the area, respectively.} \label{fig114A}
\end{figure}

Considering horizontal velocity vector fields, we investigate the correlation between the LCT velocities and the plasma
velocities coming from different heights of the simulation box. Upper panel of Fig.~\ref{figcoor} shows the correlation
coefficient as a function of depth. The zero height level is chosen at the spatial average of the level of optical depth
unity. The actual plasma velocities have been spatially averaged over the LCT tracking window size. The solid line
corresponds to a correlation coefficient of a $47$-minute average of both velocity fields, while the dotted-dashed line shows
an average of the correlation coefficients calculated individually for each of the $81$ velocity snapshots within the $47$
minutes. Clearly, time-averaging increases the correlation between the two velocity fields, indicating that the LCT
velocities are a reliable representation of the plasma velocities, in particular at timescales longer than granule lifetime
scale. We find the highest correlation coefficients for the plasma velocities coming from roughly $100$ km below the surface.
Similar results were found by Rieutord et al. (2001) using a simulation with a much coarser horizontal resolution of $95$ km.

The correlation decreases rapidly with depth in the top panel of Figure~\ref{figcoor}. This does not, however, indicate that
the LCT velocities do not reflect deeper convective structures. To investigate this we look at the correlation between the
LCT velocity divergence and the actual vertical plasma velocities as a function of depth. The results are shown in the bottom
panel of Figure~\ref{figcoor}. The vertical velocities here were spatially smoothed using a filter of the same size as the
LCT tracking window. These results indicate that the LCT algorithm is detecting the organization of convective structures
which extend at least as deep as the simulation (ie to deeper than $-1.7$~Mm below the surface). While the horizontal
velocities show less organization in the deeper layers (they are mainly resulting from mass conservation in the strongly
stratified system), the vertical velocities are strongly affected by the pattern of downflows, which stays roughly the same
from the surface to the bottom of the simulation box.
\section{Mesogranular flow patterns}
\subsection{Definition}
We define mesogranules in the simulation as patches of positive horizontal velocity divergence, analogous to the definition
used in the case of solar observations and also in the two-dimensional cellular model of Matloch et al. (2009). The analysis
procedure is identical for both the cellular model and the numerical simulation. The intensity images from the simulation
have a cadence of $30$ sec. The first step is to apply the LCT algorithm to extract the horizontal velocity field from the
displacement of granules. Then we average the resulting velocity fields over a given averaging time, $t_{a}$, and calculate
the velocity divergence. Next, mesogranules are identified as patches for which the divergence exceeds a predefined threshold
value. The level is set in the following way: for each $t_{a}$-averaged map, the rms value of the velocity divergence is
determined, and then the time average of the rms, $\Lambda$, over the whole dataset is calculated. In addition, all patches
smaller than $0.7$ of the average granule area are disregarded, to be consistent with the two-dimensional cellular model.

Individual mesogranules are tracked in time and both their lifetime and the lifetime-averaged area are determined. The
tracking algorithm works as follows: first, the mesogranules are labelled in each mesogranule image. Next, for each pair of
subsequent mesogranule images, the algorithm finds the mesogranules that show the maximum overlap in both images. Unless a
splitting has occurred, such cells are taken to be the same mesogranule. The above scheme works well because the cadence of
the divergence images is sufficiently high ($30$ sec) so that the mesogranules do not significantly shift their position
between subsequent images. The panels at the bottom of Fig.~\ref{velovect} show an example of the velocity divergence patches
(mesogranules) lying above the $0.5\Lambda$ threshold (left panel), and $0.7\Lambda$ threshold (right panel) for the
averaging time $t_{a}=60$ min. %
\begin{figure}[h]
\begin{center}
\includegraphics[width=85mm]{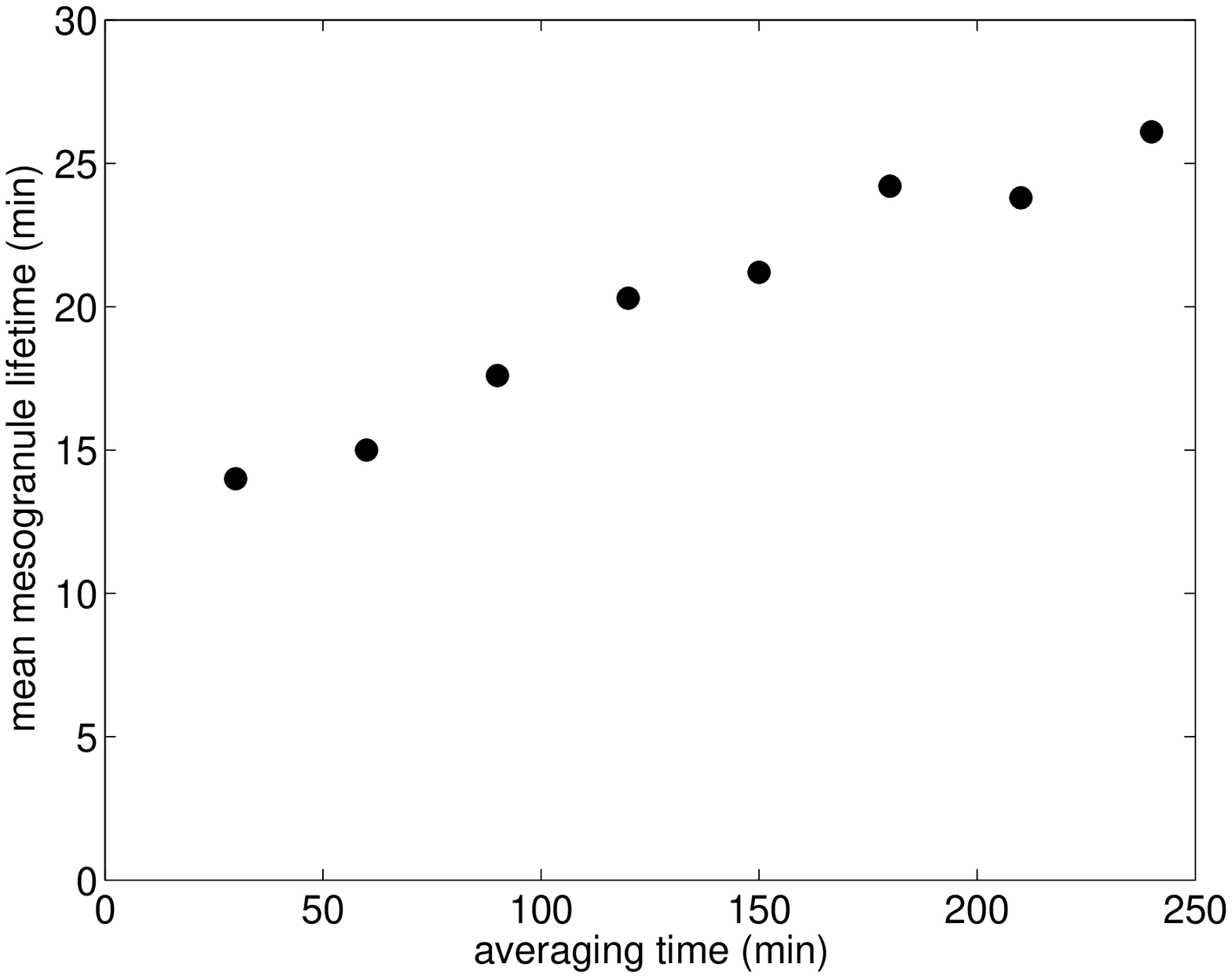}
\includegraphics[width=85mm]{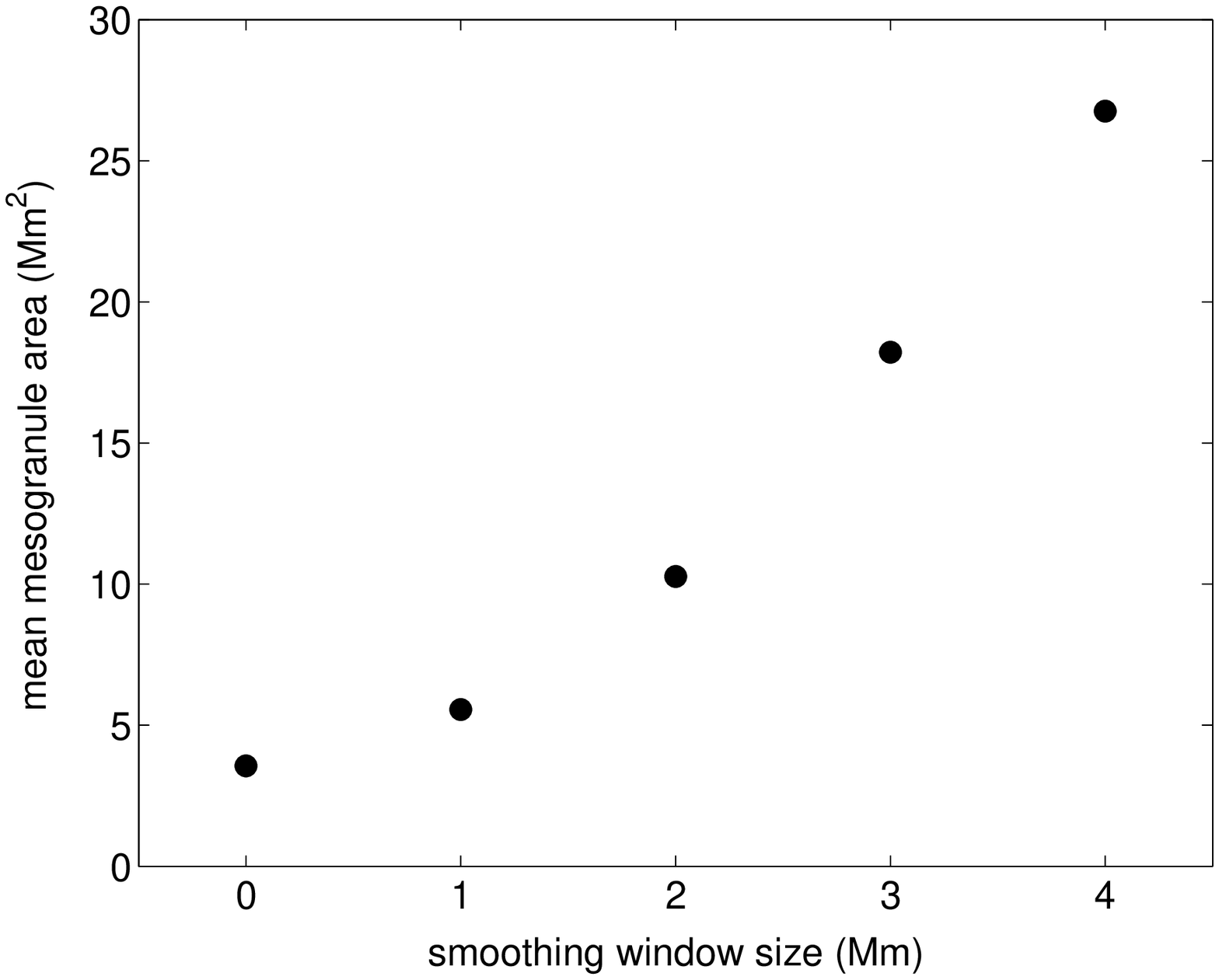}
\end{center}
\caption{Dependence of the mean mesogranule lifetime on the averaging time (top) and of mean mesogranule
area on the spatial smoothing window size (bottom).} \label{guy}
\end{figure}
\subsection{Properties}
Figure~\ref{fig114A} shows the mesogranule statistics obtained for a threshold level of $0.5\Lambda$ and an averaging time
$t_{a}=60$ min. Both histograms are well approximated by power laws, with exponents $-1.2$ for lifetime and $-2.4$ for the
area, similar to the result obtained from the two-dimensional cellular model, where exponents $-1.3$ and $-3.4$ were found,
respectively. (see Fig.~17 in Matloch et al. 2009). The power-law behaviour of the distributions suggests that no
characteristic scales are associated with the pattern.

We investigate how the mean mesogranule properties depend on the analysis parameters: the averaging time $t_{a}$, the spatial smoothing window
size and the divergence threshold level. The LCT procedure introduces spatial smoothing due to the LCT tracking window, which has to be of the
size of the tracers (here granules). Hence, the flow field obtained by LCT is effectively filtered from all contributions below a spatial
scale of roughly $2.5$ Mm (Rieutord et al. 2001, 2010). We take the actual plasma velocities, and apply spatial smoothing windows of different
sizes, as well as different averaging times, in order to study how it influences the resulting mesogranular pattern. Figure~\ref{guy} shows the
dependence of the mean mesogranule lifetime on the averaging time, $t_{a}$ (upper panel), and the dependence of the mean mesogranule area on the
smoothing window size (lower panel). The lifetime increases roughly linearly with the averaging time, while the area increases as a square of the
spatial smoothing window size. We also find that the mean mesogranule area does not depend on the averaging time. These results are similar to
those from the cellular model (see Fig.~18 in Matloch et al. 2009), and imply that the mesogranular pattern has no characteristic scale.

\begin{figure}[h]
\begin{center}
\includegraphics[width=90mm]{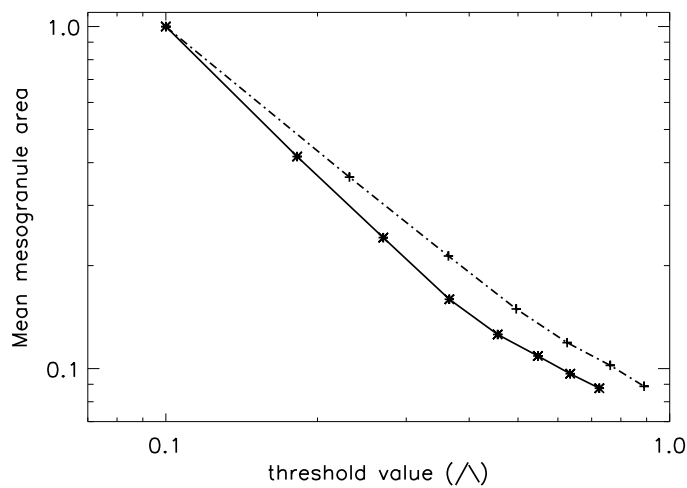}
\end{center}
\caption{Dependance of the mean mesogranule area on the threshold value for the cellular model (solid line) and the numerical
simulation (dashed-dotted line), obtained for an averaging time of $60$ min. The values of the mesogranule area have been
normalized so that the area equals $1$ for a threshold of $0.1\Lambda$ for both models.} \label{fig113}
\end{figure}

So far we have seen that the velocity divergence field and the corresponding mesogranular pattern produced in the
numerical simulation are similar to that resulting from the cellular model (Matloch et al. 2009). To further investigate the
structure of the divergence fields in both models, we plot in Fig.~\ref{fig113} the dependence of the average mesogranule
size (obtained for a fixed averaging time $t_{a}=1$ hour) on the threshold value, $\Lambda$, for mesogranule definition. The
values of the mesogranule area have been normalized so that the area equals $1$ for a threshold of $0.1\Lambda$ for both
models.
The behaviour of the curves in Fig.~\ref{fig113} is similar: they can be approximated with a power law with exponents $-1.5$
and $-1.1$ for the cellular model and numerical simulation, respectively. This supports the conclusion that the structure of
the velocity divergence field on scales larger than granulation produced by the cellular model is very similar to that in the
MURaM simulation.

\section{Conclusions}

We found mesogranular structures in a shallow (bottom $\sim1.7$ Mm below $\tau=1$ level) hydrodynamical simulation of solar
surface convection. The mesogranules were found to have no intrinsic temporal or spatial scales and showed a power-law
distribution of sizes and lifetimes. The mean values depended on the averaging times and the size of the spatial smoothing
window used in the analysis. The LCT velocity was found to be correlated with real convective motions: horizontal velocities
near the surface and vertical velocities across a broad height range.

The properties of mesogranular flow patterns emerging in the numerical simulation correspond very well to those in the
cellular model (Matloch et al. 2009) when the same analysis methods are used in both models. The distributions of mesogranule
areas and lifetimes, as well as the dependence of the mean values of the mesogranule area and lifetime on the analysis
parameters, follow the same laws. This suggests that the mesogranular flows do not represent a distinct convective scale.


\begin{thebibliography}{}
%
\bibitem[Cardena et al.(2003)]{Cardena}Dom\'{i}nguez Carde\~{n}a, I. 2003, A\&A, 412, L65
\bibitem[Cattaneo et al.(2001)]{Cattaneo2001}Cattaneo, F., Lenz, D., \& Weiss, N. 2001, ApJ, 563, L91
\bibitem[Georgobiani et al.(2007)]{Georgobiani}Georgobiani, D., Zhao, J., Kosovichev, A. G., Benson, D., Stein, R. F., \& Nordlund, \AA. 2007, ApJ, 657, 1157
\bibitem[Leitzinger et al.(2005)]{Leitzinger2005}Leitzinger, M., Brandt, P. N., Hanslmeier, A., P\"{o}tzi, W., \& Hirzberger, J. 2005, A\&A, 444, 245
\bibitem[Matloch et al.(2009)]{Matloch}Matloch, L., Cameron, R.,Schmitt, D., \& Sch\"{u}ssler, M. 2009, A\&A, 504, 1041M
\bibitem[Nordlund (2005)]{nord}Nordlund, $\AA$. 1985, Solar Phys., 100, 209
\bibitem[Nordlund et al.(2009)]{Nordlund}Nordlund, $\AA$, Stein, R. F., \& Asplund, M. 2009, Liv. Rev. in Sol. Phys., 6, 2
\bibitem[November et al.(1981)]{November}November, L. J., Toomre, J., Gebbie, K. B., \& Simon, G. W. 1981, ApJ, 245, L123
\bibitem[Ploner et al.(1999)]{PlonerA}Ploner, S. R. O., Solanki, S. K., \& Gadun, A. S. 1999, A\&A, 352, 679
\bibitem[Ploner et al.(2000)]{PlonerAs}Ploner, S. R. O., Solanki, S. K., \& Gadun, A. S. 2000, A\&A, 356, 1050
\bibitem[Rieutord et al.(2001)]{Rieutorda}Rieutord, M., Roudier, T., Ludwig, H.-G., Nordlund, \AA., \& Stein, R. 2001, A\&A, 377, L14
\bibitem[Rieutord et al.(2010)]{Rieutorda}Rieutord, M., Roudier, T., Rincon, F., Malherbe, J. M., Meunier, N., Berger, T., Frank, Z., 2010, A\&A, 512A,
4R
\bibitem[Roudier et al.(1998)]{Roudier1998}Roudier, Th., Malherbe, J. M., Vigneau, J., \& Pfeiffer, B. 1998, A\&A, 330, 1136
\bibitem[Roudier et al.(2003)]{Roudier2003}Roudier, Th., Ligni\`{e}res, F., Rieutord, M., Brandt, P. N., \& Malherbe, J. M. 2003, A\&A, 409, 299
\bibitem[Simon et al.(1991)]{Simon SA}Simon, G. W., Title, A. M., \& Weiss, N. O. 1991, ApJ, 375, 775
\bibitem[Stein et al.(1998)]{stein1998}Stein , R. F., \& Nordlund, \AA. 1998, ApJ, 499, 914
\bibitem[V\"{o}gler (2003)]{VoglerS}V\"{o}gler, A., 2003, \textit{PhD Thesis}, Univ. G\"{o}ttingen,
http://webdoc.sub.gwdg,de/diss/2004/voegler/
\bibitem[V\"{o}gler et al.(2005)]{Vogler}V\"{o}gler, A., Shelyag, S., Sch\"{u}ssler, M., Cattaneo, F., Emonet, T., \& Linde,T. 2005, A\&A, 429, 335
\bibitem[Welsh et al.(2004)]{Welsh2004}Welsh, B. T., Fisher, G. H., \& Abbett, W. P., 2004, ApJ, 610, 1148
%
\end{thebibliography}
\end{document}